# Bidirectional Self-Folding with Atomic Layer Deposition Nanofilms for Microscale Origami

*Baris Bircan[†], Marc Z. Miskin[§,‡,#], Robert J. Lang[∥], Michael C. Cao[†], Kyle J. Dorsey[†], Muhammad G. Salim[⊥], Wei Wang[§], David A. Muller[†,‡], Paul L. McEuen[§,‡], Itai Cohen*[,§,‡]*

[†]School of Applied and Engineering Physics, 271 Clark Hall, Cornell University, Ithaca, New York 14853, United States

[§]Laboratory of Atomic and Solid State Physics, 511 Clark Hall, Cornell University, Ithaca, New York 14853, United States

[‡]Kavli Institute at Cornell for Nanoscale Science, 420 Physical Sciences Building, Cornell University, Ithaca, New York 14853, United States

[∥]Robert J. Lang Origami, Alamo, California 94507, United States

[⊥]Cornell Center for Materials Research, 627 Clark Hall, Cornell University, Ithaca, New York 14853, United States

**ABSTRACT**

*Origami design principles are scale invariant and enable direct miniaturization of origami structures provided the sheets used for folding have equal thickness to length ratios. Recently, seminal steps have been taken to fabricate microscale origami using unidirectionally actuated sheets with nanoscale thickness. Here, we extend the full power of origami-inspired fabrication to nanoscale sheets by engineering bidirectional folding with 4 nanometer thick atomic layer deposition (ALD) $SiN_x$ - $SiO_2$ bilayer films. Strain differentials within these bilayers result in*

*bending, producing microscopic radii of curvature. We lithographically pattern these bilayers and localize the bending using rigid panels to fabricate a variety of complex micro-origami devices. Upon release, these devices self-fold according to prescribed patterns. Our approach combines planar semiconductor microfabrication methods with computerized origami design, making it easy to fabricate and deploy such microstructures en masse. These devices represent an important step forward in the fabrication and assembly of deployable micromechanical systems that can interact with and manipulate micro and nanoscale environments.*

**TEXT**

Due to their scale invariance, origami design principles have been used to create complex systems across various sizes, such as deployable solar panels on spacecraft,[1,2] centimeter scale programmable materials[3,4] and robots,[5,6] and microelectromechanical systems (MEMS).[7] At the microscale, origami has proven to be especially advantageous, as it has allowed the use of planar lithographic fabrication methods to build 3D structures that remain inaccessible to other manufacturing processes. Within the last decade, folding at the microscale has been demonstrated using metallic thin films,[8,9] polymers[10,11] and atomically thin films of hard materials.[12,13,14,15]

As a result of their high Young's modulus and low bending stiffness, ultra-thin films of hard materials are an excellent material choice for self-folding micro-origami. When fabricated at nanoscale thicknesses, these films can repeatably and elastically deform to micron scale radii of curvature,[16] while producing sufficient force output to lift rigid panels 1000 times their thickness.[17] Atomic layer deposition (ALD) enables the fabrication of oxide and nitride films with nanometer thickness, high mechanical integrity, high uniformity and low pinhole defect density.[18,19,20,21] Thus, this technology opens the door to a new generation of atomically thin sheets that define the smallest possible size scale for self-folding. Recent work has demonstrated ALD based, 2 nm thick unidirectional bending actuators to build elementary self-folding microstructures like tetrahedra and cubes.[14]

In general, the algorithmic mapping of a 3D shape to a fold pattern requires the assignment of fold angles ranging from -180° to +180°.[22,23] Therefore, unidirectional fold actuation limits the space of shapes that can be fabricated to relatively simple ones. Bidirectional folding is required to create combinations of mountain (downward) and valley (upward) folds found in complex origami designs (Fig. 1A).[24] While inorganic ALD films have emerged as a promising class of materials for creating microscale folds, systems based on these materials have not yet demonstrated bidirectional folding for origami-based 3D self-assembly.[25]

Here, we present 4 nm thick, ALD oxide - nitride bilayer films that achieve bidirectional bending, and we use these films to develop a scalable microfabrication process that can be generalized to create rigidly foldable[26] origami. The ultra-thin ALD bilayers at the center of our approach convert an internal strain mismatch into a bending response. Provided that a certain bilayer stack of ALD films bends in one direction, the stacking order can be inverted to produce bending in the opposite direction. In order to create bidirectional folding, we combine ALD bilayers that bend in opposite directions and add rigid panels to localize bending (Fig. 1B). Our fabrication process is based entirely on conventional semiconductor manufacturing techniques and lets us take advantage of lithographic methods to pattern and release our microscale origami devices in parallel. When released into solution, these devices self-fold into configurations defined by prescribed fold patterns. Using this approach, we fabricate a variety of complex micro-origami devices, which can withstand highly corrosive acidic environments and processing temperatures up to 150 °C. The development of these self-folding nanofilms, together with a process that can be generalized to create arbitrarily complex origami, extends the full power of origami-inspired fabrication to nanoscale sheets.

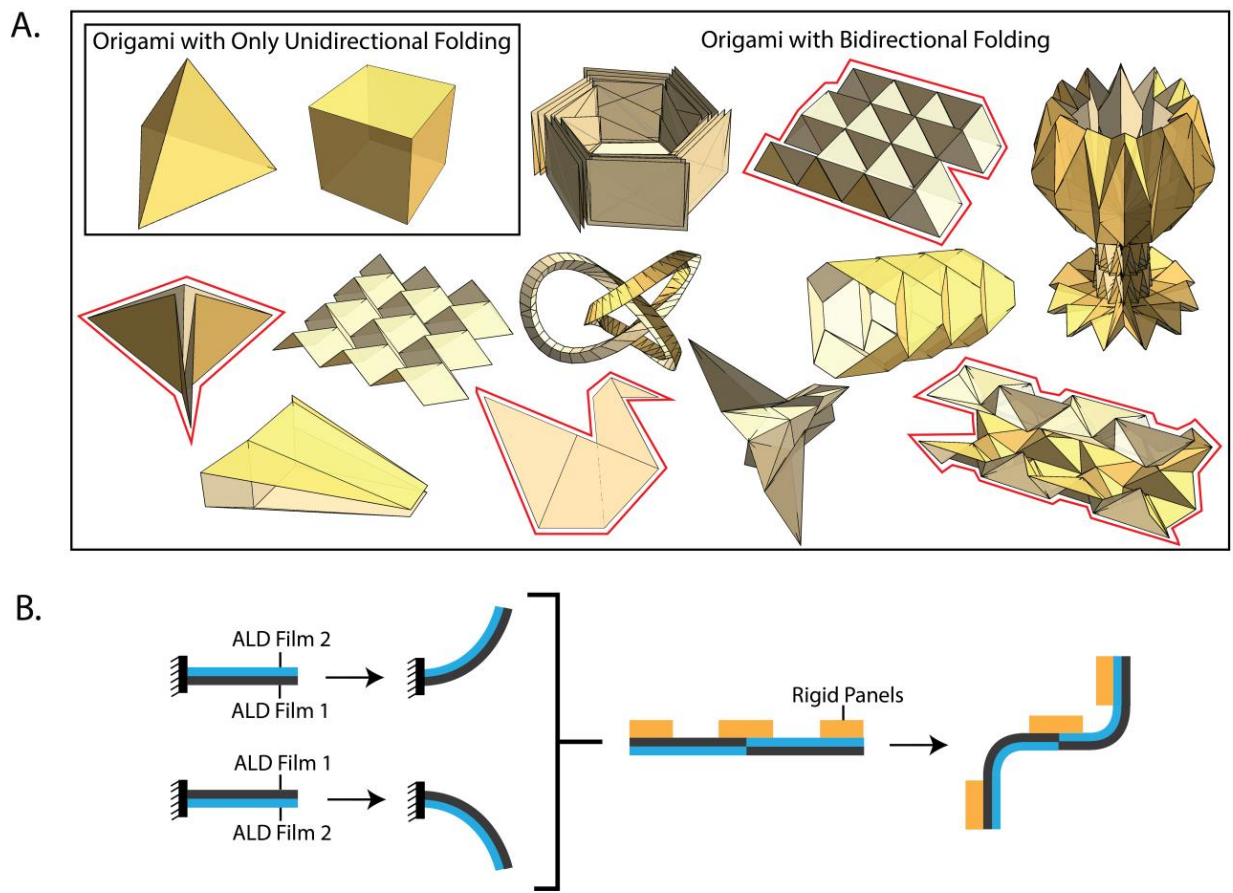

**Figure 1.** Bidirectional folding with atomic layer deposition (ALD) bilayer films. Mapping of an arbitrary 3D shape to an origami fold pattern requires the assignment of fold angles ranging from -180° to +180°. As a result, unidirectional fold actuation limits the space of shapes that can be fabricated to relatively simple ones (A). Bidirectional folding is required to create combinations of mountain (downward) and valley (upward) folds found in complex origami designs. We choose to experiment with the four designs highlighted in red to demonstrate our approach can create microscale origami of variable complexity. To engineer self-folding at the microscale, we use ALD bilayer films, which convert an internal strain mismatch into a bending response (B). If a certain bilayer stack of ALD films produces bending in one direction, the stacking order can be inverted to create bending in the opposite direction. We create bidirectional folding by combining sheets that bend in different directions and adding rigid panels to localize bending.

We identify ALD $SiN_x$ and $SiO_2$ as suitable materials for use in our micro-origami devices due to their low growth stresses and similar Young's moduli.[27] To obtain the smallest possible sheet thickness and maximum bending deformation, while ensuring that fully nucleated ALD films can be reproducibly deposited, we fabricate bilayers consisting of 2 nm of $SiO_2$ and 2 nm of $SiN_x$. Figure 2A (and Supporting Video 1) show a hinge made with a $SiN_x$ (bottom) - $SiO_2$ (top) bilayer and a flat panel made of SU-8 polymer. When released in solution, this hinge folds upwards, thus identifying the film stacking order that will be used for valley (upward) folding.

Residual stresses form within the $SiN_x$ and $SiO_2$ films due to the nucleation and growth processes that occur during atomic layer deposition; and thermal stresses form due to the elevated growth temperatures and the mismatch between the thermal expansion coefficients of the films. A combination of these stress components gives rise to the intrinsic curvature of the bilayer, which we localize to create folds. We measure this curvature to be 0.1 $\mu m^{-1}$ for folding in both directions, which is consistent with a strain on the order of 0.01% given the thickness of the bilayer (details on bilayer curvature characterization can be found in SI). Since the strains present are two orders of magnitude below the fracture strains for silicon nitride and silicon dioxide,[28,29] the hinge operates elastically, demonstrating full recovery after being deformed by an applied force. Electron energy loss spectroscopy (EELS) (Fig. 2B) and X-Ray Photoelectron Spectroscopy (XPS) (Fig. 2C) on the ALD film stack used in this hinge confirm that the bilayer consists of approximately 2 nm thick layers of $SiN_x$ and $SiO_2$ (details on ALD film stack characterization can be found in SI.).

Combined with knowledge of the ALD bilayer's intrinsic curvature, the ability to invert the bending direction by changing the process order makes the fabrication of mountain (downward) and valley (upward) folds on separate substrates relatively straightforward.

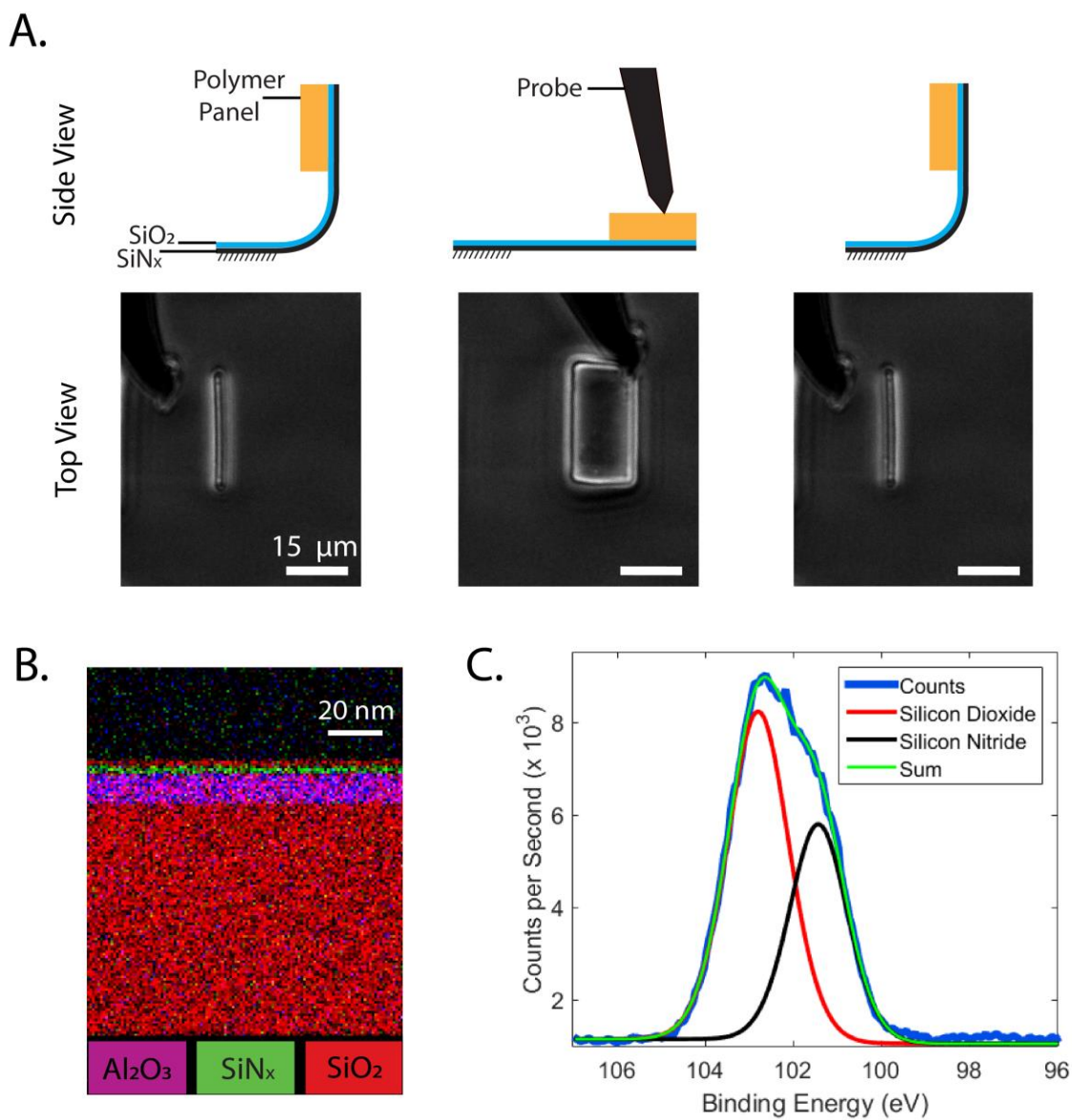

**Figure 2.** Operation and structure of an ALD $SiN_x$ - $SiO_2$ hinge. A hinge made with an ALD $SiN_x$ (bottom) - $SiO_2$ (top) bilayer and a flat SU-8 panel folds upwards when released (A), identifying the device structure needed for valley folding. The bilayer hinge operates elastically, demonstrating full recovery after being deformed by a micromanipulator probe. EELS analysis (B) of the ALD bilayer confirms that the ALD nitride and oxide films making up the bilayer meet their target thickness of 2 nm. Elemental signals for oxygen (red), aluminum (blue) and nitrogen (green) have been isolated to create the color-coded EELS map. The bilayer is seen on top of a 10 nm thick layer of $Al_2O_3$, which is colored pink due to a combination of blue (aluminum) and red (oxygen). (C) High resolution XPS scans of the ALD film stack further confirm the internal structure of the bilayer. Photoelectron counts from the film stack (blue) are fit well by the sum (green) of two Voigt profiles centered at Si 2p binding energy values corresponding to $SiO_2$ (red) and $SiN_x$ (black) (details on ALD film stack characterization can be found in SI).
6

However, the similar etch behavior of the two ultra-thin ALD films presents a challenge for fabricating complex patterns of alternating fold direction on a microscopic device. During fabrication, sequential photolithography and etching steps risk damaging or completely removing previously patterned parts of each device. Here, we develop a fabrication process that incorporates an aluminum etch mask to protect device layers during processing without sacrificing geometrical complexity. This approach allows us to reliably create combined bidirectional folds, where fold angles can range from -180° to +180°.

The process flow needed to fabricate a general microscopic origami shape is outlined in Figure 3. We start by thermally evaporating an aluminum release layer onto a substrate. Then, we grow an ALD $SiO_2$ - $SiN_x$ film stack. Next, we thermally evaporate another layer of aluminum to be used as an etch mask in subsequent steps (Fig. 3A). We spin coat the substrate with a standard positive tone photoresist and use contact photolithography to define the mountain (downward) fold regions of the origami shape. We then perform plasma etching to transfer the photoresist pattern into the aluminum etch mask and the ALD stack underneath it (Fig. 3B-3C). We remove the photoresist by soaking the substrate in solvent overnight.

Next, we deposit an inverted $SiN_x$ - $SiO_2$ ALD film stack (Fig. 3D). The conformal nature of the ALD process results in a fully coated substrate, which ensures the ALD films making up each device are continuous. We spin coat the substrate with positive photoresist and use contact photolithography followed by plasma etching to pattern the inverted bilayer film (Fig. 3E-3F). This step defines the ALD hinges that make up the valley (upward) folds as well as the regions where the flat panels will be placed. During this patterning step, the mountain (downward) fold regions are protected from the plasma by the aluminum mask. We leave a 3-5 μm overlap between the mountain and valley photoresist patterns to ensure the ALD sheets making up each device are continuous. After pattern transfer, we remove the photoresist by soaking the substrate in solvent overnight.

Next, we spin coat the wafer with SU-8 photoresist and use contact photolithography to pattern 1 μm thick flat panels (Fig. 3G). Using the measured thicknesses and known elastic constants of our materials,[17, 27, 30] we estimate that the bending stiffness $Yt^3/[12(1-\nu^2)]$ ($Y$: Young's modulus, $t$: thickness, $\nu$: Poisson's ratio) of the SU-8 panels is five orders of magnitude larger than that of the self-folding ALD sheets. Therefore, the SU-8 panels are effectively rigid compared to the ALD hinges and can be used to define flat facets for each origami shape.

Lastly, we use an HCl wet etch to dissolve the aluminum release layer and etch mask to release the origami devices into solution, where they freely fold. All device layers are designed to have aligned holes exposing the release layer to the etch solution, so that undercut time during the release process is reduced.

Our fabrication approach is applicable to any rigidly foldable origami pattern that folds simultaneously, and allows us to design and construct a large variety of structures. We use an extended version of the Tessellatica[31] package to automate the lithography mask design process. Through this software package, fold patterns modeled in Tessellatica are converted into design files for a set of three lithography masks. We then directly incorporate each mask into the corresponding lithography step in our process to microfabricate the fold patterns. The Tessellatica package used in generating our lithography masks can be found under reference 32.

The software we use to create our lithography masks converts fold lines of infinitesimal width into hinges with finite width. The direction for each fold is determined by the ALD bilayer stack order at the fold location, and the magnitude of the fold angle is given by the ratio of the exposed length between rigid panels to the intrinsic radius of curvature produced by the bilayer.

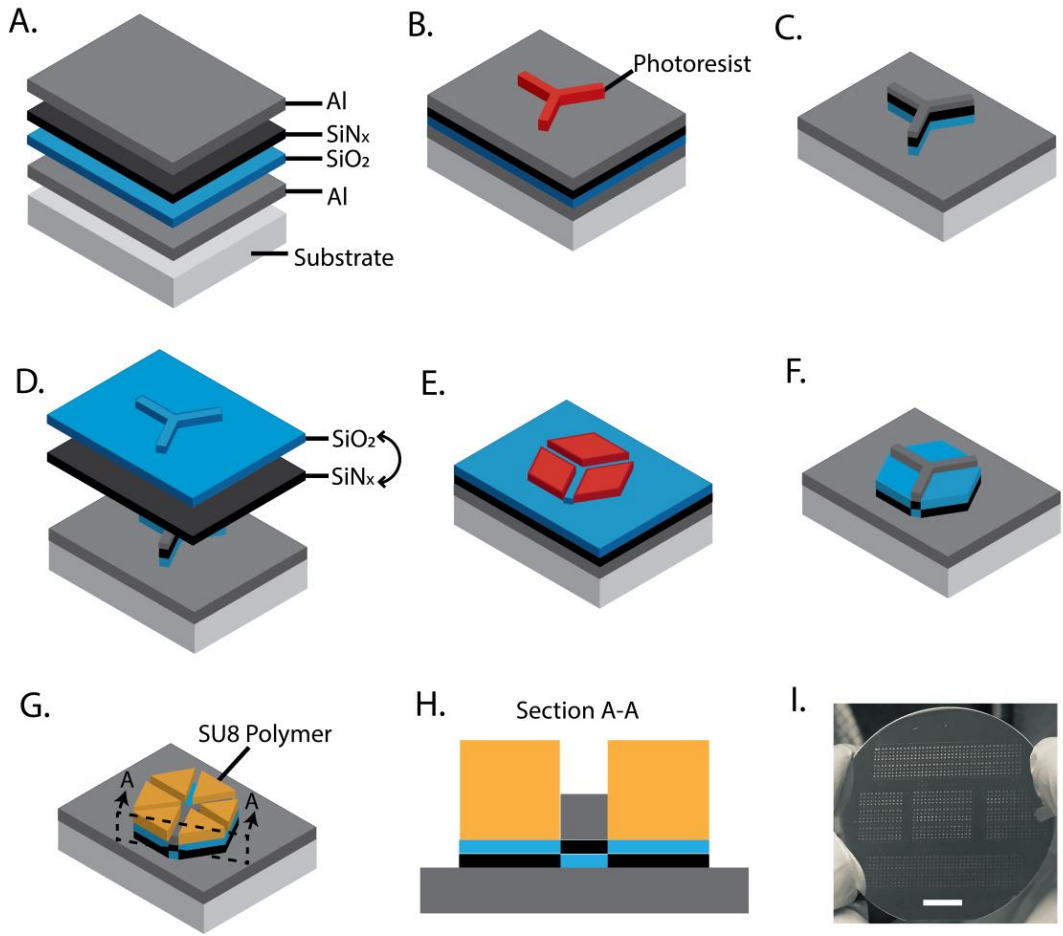

**Figure 3.** Process flow to fabricate self-folding micro-origami with ALD nanofilms. Our process can be used to microfabricate any rigidly foldable origami pattern that folds simultaneously. The steps required to fabricate a micro-origami device are described using the origami hexagon as an example: (A)Deposit an aluminum release layer on a transparent substrate, followed by an ALD $SiO_2$ - $SiN_x$ film stack and another layer of aluminum for etch masking. (B)Use photolithography to define the mountain (downward) fold regions of the origami shape. (C)Perform plasma etching and transfer the photoresist pattern into the aluminum etch mask and the ALD stack underneath it. (D)Deposit an inverted $SiN_x$ - $SiO_2$ ALD film stack and coat the entire substrate. (E)Use photolithography to define the ALD hinges that make up the valley (upward) folds as well as the regions where the flat panels will be placed. (F)Use plasma etching to pattern the inverted ALD bilayer, while the aluminum mask protects the previously patterned mountain fold regions. (G)Use photolithography to pattern 1 μm thick flat SU-8 panels. (H)Device cross section after processing is completed. (I)Unreleased origami devices on a processed two-inch wafer are seen as an array of points to the naked eye. Scale bar is 1 cm. Immersing the wafer in an HCl solution removes the aluminum release layer and etch mask, allowing the devices to deploy and self-fold.

Microscopic origami devices of variable complexity made using this approach are shown in figure 4. In each row, from left to right, we show the fold pattern for an origami shape, a render of the target 3D shape, a finite element model of the self-folding device, the micro-origami device attached to the substrate in its flat state, and the micro-origami device in its folded state. We image our devices with an optical microscope and create the images of the folded devices by taking a 2D projection of images collected from different focal planes. For devices with larger panels, such as the ones in figure 4A and 4B, we use image slices obtained with a confocal microscope to create the projections to better capture the 3D geometries. Since the self-folding ALD sheets we develop here are transparent, only the flat SU-8 panels on each device are visible in the micrographs. Knowing that the SU-8 panels are connected by transparent ALD hinges of finite length in our finished devices, we infer the overall 3D shapes from the relative positions of the panels. We include finite element models of the devices together with their micrographs to visualize the overall 3D shapes by highlighting the ALD sheets, which are otherwise not visible.

The simplest structure we demonstrate is the origami hexagon (Fig. 4A), which has six folds. The apparent discontinuity of the bottom pair of edges in the folded device is an artifact of the transparency of the ALD sheets, since the two adjacent SU-8 panels are visible, but the transparent curved hinge connecting them is not.

As a more complex example, we show the rigidly foldable duck (Fig. 4B-17 folds), which is less than half the size of the current state-of-the-art micro-origami bird.[11] Due to the increased density of folds near the head of the duck, the microscopic device design we obtain from our software consists only of ALD bilayers near the head of the duck, which makes this region transparent. Since we can verify correct kinematic positioning of all the SU-8 panels, we conclude that the device has self-assembled as designed.

Lastly, we demonstrate that our fabrication process can also be used to build structures that are an order of magnitude more complex, such as the spacer Miura-ori (Fig. 4C-71 folds), and the waterbomb (Fig. 4D-120 folds) tessellation, which has the same diameter as a human hair. The folded form of the spacer Miura-ori device has SU-8 panels tilting alternatingly upward and downward, separated by fully flat-folded regions, as designed. Due to the high density of folds in the waterbomb tessellation, the conversion of this origami design into a microscopic device requires that most of the structure consists of only ALD bilayers. Even though the visible SU-8 panels constitute only a small fraction of the overall area, we can verify that the distribution of folded regions gives rise to a cylindrical folded form with the correct aspect ratio and conclude that the device has self-assembled as designed. Our technique thus demonstrates the ability to design and fabricate microscale origami ranging from relatively simple shapes with less than 10 folds to complicated tessellations with more than 100 folds.

When process limitations are considered, stiction is the most important factor affecting our device yield. During the wet release process, adhesion between the substrate and the origami devices can lead to partial release of devices, which prevents full deployment and assembly into the correct target geometries. Similarly, the self-folding ALD sheets can stick to themselves upon contact, resulting in devices that get stuck in a misfolded position, preventing assembly into the correct target geometries. We introduce surfactant into the solution environment to reduce these impacts of stiction. The resulting device yield varies from about 50% for simple geometries like the hexagon to less than 10% for complex geometries like the spacer Miura-ori. To minimize fabrication costs, we dice our substrates into 2 cm x 2 cm pieces containing about a thousand devices each, and we experiment on one piece at a time. The substrate and die sizes are not limited by any of our fabrication steps, and the process can easily be scaled up by using larger substrates.

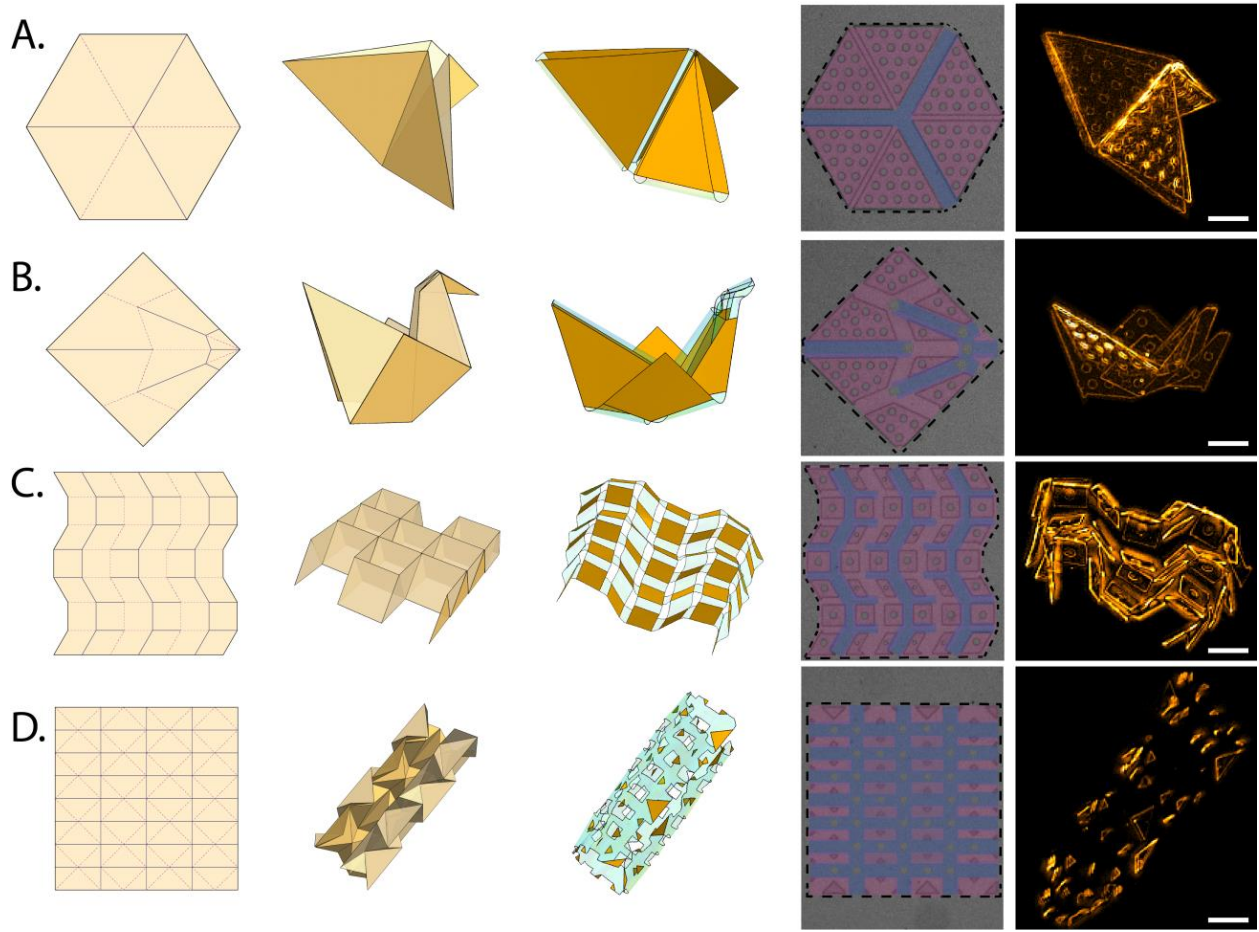

**Figure 4.** Self-folding micro-origami made with ALD nanofilms. ALD $SiN_x$ - $SiO_2$ bilayer films can be used to fabricate microscale origami of variable complexity. From top to bottom, the examples we show here are: the origami hexagon (A) with 6 folds, the rigidly foldable duck (B) with 17 folds, the spacer Miura-ori (C) with 71 folds, and the waterbomb tessellation (D) with 120 folds. The leftmost column of the figure shows fold patterns corresponding to each origami shape, with mountain (downward) fold lines drawn in blue and valley (upward) fold lines drawn in pink. The second to left column shows renders of each folded origami shape. The middle column shows finite element models of the expected folded device configurations, with rigid panels represented in orange and the transparent self-folding sheets represented in light blue. We note that the finite element models include holes in their vertices to avoid singularities in computation (model details can be found in SI). The second to right column shows microscale origami devices attached to the substrate in their flat state, with dashed lines outlining each device. The transparent ALD sheets making up the devices are visualized using false color. Regions with $SiO_2$ (bottom) - $SiN_x$ (top) bilayers are shaded blue and regions with $SiN_x$ (bottom) - $SiO_2$ (top) bilayers are shaded pink. Vertices that have had material removed from them are left uncolored. The grid of holes evident in the SU-8 panels is introduced into all the device layers in order to expose the release layer to the etch solution and reduce the time needed to undercut the devices during the wet release process. The rightmost column shows microscale origami devices after they have been released and allowed to self-assemble into their folded configurations, with the SU-8 panels colored orange. Scale bars are 50 μm.

Overall, our approach establishes a complete toolkit for origami-inspired microfabrication with hard materials: using software tailored for our ALD sheets, any fold pattern can be converted into a set of lithography masks, which can be directly included into our process to mass fabricate the designs at the microscale. This process opens the door to a whole class of deployable origami-inspired microscale mechanisms and machines that can explore and interact with their environment.

The use of ultra-thin ALD bilayers for origami-inspired microfabrication establishes a strong foundation for future work. The process we have developed in this work can easily be generalized to include bilayers and trilayers made up of any ALD material, as long as the individual ALD films have sufficiently small internal stress gradients so that the only significant curvature formed is due to the strain mismatch between the two layers. The wide material palate offered by ALD[33], combined with surface functionalization methods, provides a way to incorporate various capabilities into our devices, such as voltage, chemistry, temperature and biomolecule-based sensing and actuation. Stepwise fold actuation mechanisms can be introduced to enable sequential folding, allowing more reliable folding pathways[34,35] and even more complicated origami structures. Since our fabrication process is entirely carried out in a cleanroom using semiconductor processing tools, it can also be extended to include on-board microelectronics and/or magnets[17,36] to control and actuate our self-assembled structures. If there is need to remove the devices from liquid and into atmosphere, critical point drying can be employed to avoid damage due to surface tension forces. The integration of these capabilities will make our micro-origami devices a powerful tool for interacting with and manipulating environments at the micrometer and nanometer scales.

AUTHOR INFORMATION

**Corresponding Author**

*E-mail: Itai Cohen, ic64@cornell.edu

**Present Addresses**

#Department of Electrical and Systems Engineering, 203 Moore Building, University of Pennsylvania, Philadelphia, Pennsylvania 19104, United States

**Author Contributions**

B.B., M.Z.M., P.L.M. and I.C. conceived the experiments. B.B. fabricated the samples and performed the experiments with assistance from M.Z.M. and under P.L.M.'s and I.C. 's supervision. R.J.L. contributed origami fold patterns and developed software for automated lithography mask design. M.C.C. and K.J.D. performed cross sectional imaging and thickness characterization of ALD bilayers under D.A.M.'s supervision. M.G.S. performed XPS analysis of ALD bilayers. W.W. created finite element models of origami shapes. B.B. and I.C. wrote the manuscript with consultation from all authors.

**Funding Sources**

This work made use of the Cornell Center for Materials Research Shared Facilities which are supported through the NSF MRSEC program (DMR-1719875). This work was supported by the U. S. Army Research Office (ARO W911NF-18-1-0032), the Cornell Center for Materials

Research with funding from the NSF MRSEC program (DMR-1719875), the Watt W. Webb Fellowship at Kavli Institute at Cornell for Nanoscale Science, and performed in part at the Cornell NanoScale Science & Technology Facility (CNF), a member of the National Nanotechnology Coordinated Infrastructure (NNCI), which is supported by the National Science Foundation (Grant NNCI-1542081).

**Notes**

Authors declare no competing interests.

ACKNOWLEDGMENTS

We thank Michael Reynolds, Tanner Pearson, Alejandro Cortese, Samantha Norris, Qingkun Liu, Meera Ramaswamy and Samuel Whitehead for insightful discussions.

ABBREVIATIONS

ALD, atomic layer deposition; EELS, electron energy loss spectroscopy; XPS, X-Ray Photoelectron Spectroscopy.

# Supporting Information for:

# Bidirectional Self-Folding with Atomic Layer Deposition Nanofilms for Microscale Origami

*Baris Bircan[†], Marc Z. Miskin[§, ‡, #], Robert J. Lang[∥], Michael C. Cao[†], Kyle J. Dorsey[†], Muhammad G. Salim[⊥], Wei Wang[§], David A. Muller[†, ‡], Paul L. McEuen[§, ‡], Itai Cohen\*[, §, ‡]*

[†]School of Applied and Engineering Physics, 271 Clark Hall, Cornell University, Ithaca, New York 14853, United States

[§]Laboratory of Atomic and Solid State Physics, 511 Clark Hall, Cornell University, Ithaca, New York 14853, United States

[‡]Kavli Institute at Cornell for Nanoscale Science, 420 Physical Sciences Building, Cornell University, Ithaca, New York 14853, United States

[∥]Robert J. Lang Origami, Alamo, California 94507, United States

[⊥]Cornell Center for Materials Research, 627 Clark Hall, Cornell University, Ithaca, New York 14853, United States

**Contents:**

1) **Sample Fabrication Details**

2) **ALD Bilayer Characterization**

3) **Finite Element Modeling**

4) **Supporting Video Description**

## 1) Sample Fabrication Details

We thermally evaporate a 300 nm thick aluminum release layer onto a clean, 430 μm thick, double side polished sapphire wafer. We then perform plasma enhanced atomic layer deposition (PEALD) in an Oxford ALD FlexAL system to deposit 20 cycles of silicon dioxide at 200 °C followed by 120 cycles of silicon nitride at 350 °C. The number of cycles for both processes correspond to an approximate deposited film thickness of 2 nm. Tris[dimethylamino]Silane (3DMAS) and oxygen plasma are used for silicon dioxide growth. 3DMAS and argon-nitrogen plasma are used for silicon nitride growth. Next, we thermally evaporate a 150 nm thick layer of aluminum on top of the ALD stack to act as an etch mask in subsequent steps. We spin coat the wafer with P20 primer and Shipley 1813 photoresist (5000 rpm, 2000 acceleration, 30s) and soft bake for 90 seconds at 115 °C. We use contact photolithography to expose the photoresist and develop to define the mountain (downward) fold regions of each origami device. We then hard bake the photoresist at 115 °C for 5 minutes. We transfer the photoresist pattern into the aluminum etch mask (with $BCl_3/Cl_2$ plasma etching), and then the ALD film stack (with $CF_4$ plasma etching). We strip the photoresist by soaking the wafer in MICROPOSIT Remover 1165 and rinsing in isopropanol and DI water. We then oxygen plasma clean our sample and perform PEALD in inverted order, depositing 120 cycles of silicon nitride at 350 °C followed by 20 cycles of silicon dioxide at 200 °C. The conformality of the ALD process ensures that the entire wafer, including the previously patterned features, is covered by the grown films. We again spin coat P20 primer and Shipley 1813 photoresist (5000 rpm, 2000 acceleration, 30s) and soft bake for 90 seconds at 115 °C. We use contact photolithography to expose the photoresist and develop to define the valley (upward) fold regions as well as the areas the flat panels will be bonded to. We then use $CF_4$ plasma etching to transfer this pattern into the inverted ALD stack. During this step, the aluminum etch mask protects the previously patterned mountain (downward) fold regions from the CF$_4$ plasma. We leave a 3 to 5 μm overlap between the two photoresist patterns to ensure the resulting ALD sheets making up each device will be continuous after etching. We strip the photoresist by soaking in MICROPOSIT Remover 1165, and rinsing in isopropanol and DI water. Next, we spin coat the wafer with SU-8 2002 photoresist (4000 rpm, 1000 acceleration, 30s) and soft bake for 1 minute at 65 °C followed by 1 minute at 95 °C. Using contact photolithography, we expose the photoresist and define panels that will make up the flat panels on each origami structure. Next, we post exposure bake the resist for 1 minute at 65 °C, followed by 2 minutes at 95 °C. We develop in SU-8 Developer for 1 minute and rinse in isopropanol. We then anneal the panels on a hot plate by heating the wafer from 95 °C to 150 °C, holding for 5 minutes, and cooling the hot plate to room temperature. Finally, we leave the samples in hydrochloric acid solution with added surfactant (sodium dodecylbenzenesulfonate). This dissolves the aluminum release layer and etch mask and releases the origami devices into solution where they can fold. All device layers are designed to have aligned holes exposing the release layer so that undercut time during the release process is reduced.

## 2) ALD Bilayer Characterization

**Bilayer Curvature Characterization**

To measure the intrinsic curvature of the ALD bilayers, we fabricate the SiN$_x$ (bottom layer) - SiO$_2$ (top layer) cantilevers shown in Figure S1A. We observe that these cantilevers bend upwards when released, and measure the resulting curvature as 0.1 μm$^{-1}$ (Figure S1A, lower panel). We use this fixed curvature value to design and fabricate the hinge shown in figure 2 of the main text. The magnitude of the fold angle is given by the ratio of the ALD hinge width between rigid SU-8 panels to the fixed 10 µm radius of curvature produced by the bilayer. This device folds upwards to 90° as designed, confirming the feasibility of our approach to creating a fold.

Since the substrate would prevent hinges with the inverted bilayer stack from bending down, we fabricate tethered devices that consist of one upward and one downward folding hinge with equal width between rigid SU-8 panels (Fig. S1B). Since the larger SU-8 panel is observed to be horizontal after release, we conclude that the magnitude of the two fold angles are equal, which indicates that the bilayer curvature is the same in both directions. This characterization is sufficient to design and fabricate any fold with angles ranging from -180° to +180°. When creating the lithography mask designs for the shapes shown in figure 4 of the main text, we again use the fixed value for the bilayer curvature to decide the ALD hinge width for each fold, regardless of direction. The direction for each fold is determined by the bilayer stack order at the fold location. The correct assembly of these geometries demonstrates that our approach can reliably create folds in both directions, even when complex elastic effects due to the coupling of folding hinges are present.

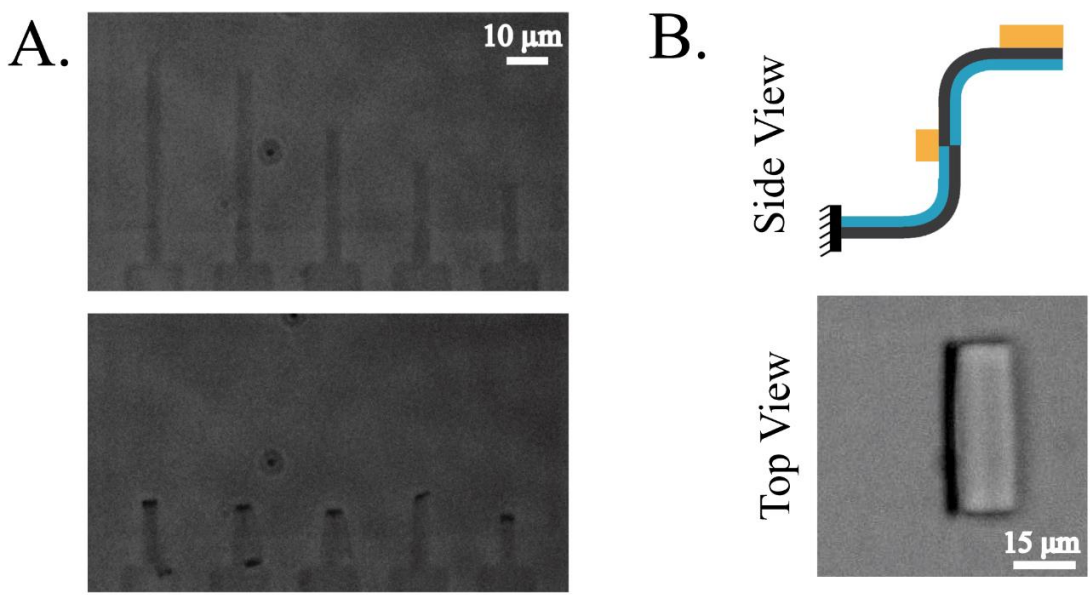

**Figure S1.** ALD Bilayer Curvature Characterization. (A) Optical micrographs of a row of five $SiN_x$ - $SiO_2$ cantilevers before (upper panel) and after (lower panel) release. We observe that these cantilevers bend upwards when released, and we measure the resulting curvature as 0.1 $\mu m^{-1}$. (B) Schematic (upper panel) and optical micrograph (lower panel) of a tethered device that consists of one upward and one downward folding hinge with equal width between rigid SU-8 panels. Since the large SU-8 panel remains horizontal after release, we conclude that the bilayer curvature is the same for folding in both directions.

**Cross Sectional STEM Imaging**

Samples for cross-sectional scanning transmission electron microscope (STEM) imaging were prepared by focused ion beam (FIB) milling and lift-out. A 500 μm thick, single side polished silicon wafer with 100 nm of thermally grown oxide was cleaned using a Piranha solution. PEALD was performed on the wafer to deposit a 10 nm thick film of aluminum oxide (at 200 °C, using trimethylaluminum and oxygen plasma), followed by 120 cycles of silicon nitride at 350 °C, and 20 cycles of silicon dioxide at 200 °C. The wafer was then cleaved along primary axes and capped with a thin layer of amorphous carbon for protection during the FIB process. Capped samples were loaded into a Thermo-Fischer Helios G4 FIB system for processing.

The FIB uses a semi-automated routine to mill out a lamella for attachment to a TEM grid. The routine begins with ion-beam assisted deposition of platinum in the area of interest, followed by patterning fiduciary marks for alignment of subsequent features. Bulk material is removed around the sample area at a beam voltage of 30 kV to produce a lamella structure for lift-out. A micromanipulator probe is inserted into the chamber and the lamella structure is attached to the probe. The probe brings the lamella to a TEM grid, attaches the lamella to the grid, and detaches the probe from the lamella. Once attached, the lamella is further thinned in a series of grazing-angle sequences while decreasing both the beam voltage and beam current. Initial thinning begins at 30 kV with a current of 0.45 nA, decreases to 16 kV and 0.12 nA, decreases further to 5 kV and 41 pA, and concludes at 2 kV and 21 pA. After thinning, the sample is removed from the FIB and stored in a vacuum desiccator. This procedure limits beam-induced sample damage and produces suitably thin samples for electron imaging and spectroscopy.

Imaging was performed on a probe-corrected Thermo Fisher Titan Themis Cryo S/TEM at 120 kV in STEM mode with a 21.4 mrad probe aperture semi-angle. The electron energy loss spectra (EELS) were collected using a Gatan Quefina dual-EELS spectrometer from 50 eV to 562 eV at 0.25 eV per channel. This energy range covers the relevant Si-$L_{2,3}$, O-K, N-K, C-K, and Al-$L_{2,3}$ edges.

**X-Ray Photoelectron Spectroscopy**

ALD bilayer samples were analyzed using a Scienta Omicron ESCA-2SR XPS with operating pressure ca. $1 \times 10^{-9}$ mBar. Monochromatic Al Kα X-rays (1486.6 eV) were generated at 300W (15kV; 20mA). Analysis spot size was 2 mm in diameter with a 0° photoemission angle and a source to analyzer angle of 54.7°. A hemispherical analyzer determined electron kinetic energy, using a pass energy of 50 eV for high resolution scans. Samples were charge neutralized using a low energy electron flood gun.

## 3) Finite Element Modeling

The finite element models of the origami devices were constructed using the finite element analysis software ABAQUS. Model dimensions were matched to the device designs used in experiments. Since the material thicknesses are significantly smaller than the in-plane dimensions, shell elements were used in modeling. The ALD sheets were modeled as composite sections consisting of silicon dioxide and silicon nitride, and the SU-8 panels were modeled as homogenous sections. All materials were assumed to be linear elastic and isotropic. In order to ensure the modeled structures would have smooth deformation fields, material was removed from the models at the vertices of the ALD sheets. Folding was simulated by prescribing isotropic expansion to the silicon nitride layer and dynamic effects were neglected. Displacement boundary conditions were applied to fix the position of a single panel in each device in the simulations.

Since the finite element models are built without material at their vertices and are subjected to simple boundary conditions that simplify computation, they are not an exact match to the experimentally realized devices. We construct these models to highlight the overall 3D shapes of our devices by visualizing the self-folding ALD sheets, which are not visible in the device micrographs.

### 4) Supporting Video Description

**Supporting Video 1:** The footage, taken using an optical microscope, shows a hinge made with an ALD $SiN_x$ (bottom) - $SiO_2$ (top) bilayer and a flat SU-8 panel. When released, the hinge folds upward at a 90° angle. A micromanipulator probe is used to apply force to the hinge and deform it into its open position. Once the probe is removed, the hinge fully recovers to its original folded position, demonstrating elastic operation.